\documentclass[12pt]{iopart}

\usepackage{iopams,bm,graphics,graphicx,epsfig,color,ulem,url}

\begin{document}

\title[SVD in parametrised tests of PN theory]{Singular value decomposition in parametrised tests of post-Newtonian theory}

\author{Archana Pai$^{1}$ and K G Arun$^{2}$}

\ead{archana@iisertvm.ac.in,kgarun@cmi.ac.in}
\address{
$^1$ Indian Institute of Science Education and Research Thiruvananthapuram, College of Engineering Trivandrum Campus, Trivandrum-695016
Kerala, India\\
$^2$ Chennai Mathematical Institute, Siruseri, 603103, India.\\
}

\begin{abstract}
Various coefficients of the 3.5 post-Newtonian (PN) phasing formula of non-spinning compact binaries moving in circular orbits is fully characterised by the two component masses. If two of these coefficients are independently measured, the masses can be estimated. Future gravitational wave observations may measure many of the 8 independent PN coefficients calculated to date. These additional measurements can be used to test the PN predictions of the underlying theory of gravity with similar PN structure. Since all of these parameters are functions of the two component masses, there is strong correlation between the parameters when treated independently. 
Using Singular Value Decomposition of the Fisher information matrix, we remove this correlations and obtain a new set of parameters which are linear combination of the original phasing coefficients. We show that the new set of parameters can be estimated with significantly improved accuracies which has implications for the ongoing efforts to implement parametrised tests of PN theory in the data analysis pipelines. 
\end{abstract}
\maketitle
\section{Introduction}\label{Intro}
Testing General Relativity (GR) and alternative theories of gravity is
one of the most important goals of directly observing gravitational wave  (GW)
signals (see \cite{SathyaSchutzLivRev09,Will05LivRev} for reviews). The Ground based GW
detectors like LIGO and Virgo are undergoing upgrades to second generation Advanced LIGO (aLIGO) and Advanced 
Virgo, making them capable of observing 1000 times the volume of the universe they were
sensitive to in the first generation. The science case studies of a third generation GW detector,
namely Einstein Telescope (ET), are being undertaken~\cite{ETScience11}. In this context, it is important to devise strategies
which enhance the capabilities of these detectors to test GR and alternative theories of gravity.
Inspiralling compact binaries are one of the most promising sources of GWs. In the adiabatic inspiral regime, the gravitational waveforms from these sources are accurately modelled using post-Newtonian (PN) approximation to GR~\cite{Bliving}. For
modelled signals, matched filtering is known to be an optimal detection
technique in Gaussian noise. Existing detection and parameter estimation
techniques are based on matched filtering. Thus, prior predictability of the
detailed nature of the waveforms becomes crucial in testing GR using inspiralling compact binaries. 

Since gravitational waveforms within alternative theories of gravity are not easy to compute, 
one tries to parametrise the possible deviations of alternative theories of gravity from GR.
There have been many proposals in the literature to test GR and its alternatives. These
proposals may be classified into two. First is to test a particular alternative theory of
gravity by bounding the value(s) of some parameters which quantifies the deviation from GR. The proposals to bound scalar-tensor theories~\cite{WillBD94,KKS95,WillYunes04,BBW05a,YagiTanaka09a,YagiTanaka09b,BertiLightScalar12} and massive graviton theories
~\cite{Will98,BBW05a,Jones05,AW09,YagiTanaka09a,YagiTanaka09b,KeppelAjith10,MYW11} are examples of the
first type. The second type is to bound the values of a set of parameters which generically captures the 
deviation from GR without referring to any particular alternative theory. Such a test was 
first proposed in \cite{BSat95} where the authors discussed the possibility of measuring the 1.5PN term in the phasing formula. This was extended to include all the PN order terms in the phasing formula in Refs~\cite{AIQS06a,AIQS06b} where the authors discussed the possibility of accurate determination of
various PN coefficients of a binary inspiral waveform from GW observations and further proposed a consistency test of these phasing coefficients in the mass plane of the binary. In Ref~\cite{MAIS10}, a similar exercise was carried
out by folding in the amplitude corrections to the PN waveforms. Recently there have been other proposals in the literature that discuss generic parametrised deviations from GR and how they affect GW detection and
parameter estimation. Refs.~\cite{YunesPretorius09,PPE2011,PPEpol2012} discussed parametrised post-Einsteinian (PPE)
bounds on alternative theories of gravity while Ref.~\cite{LiEtal2011} discussed implementing a variant of the test proposed in \cite{AIQS06b} in the LIGO data analysis context using Bayesian methods. The PPE can be viewed as a generalisation of PTPN to allow for the fact many modified theories of gravity do not possess the same PN structure as GR. In other words, the PPE tests reduce exactly to the PTPN in the limit where the former's exponent parameters (of the frequency) are taken to be same as those of PN series in GR. The effect of spin precession on determining the various PN parameters was studied in Ref.~\cite{HKJ11}. Ref.~\cite{Arun2012} discussed possible bounds on generic dipolar gravitational radiation from the gravitational wave observations. 

This work can be seen as an extension of the idea proposed in Ref.~\cite{AIQS06a} where the gravitational waveform was parametrised in terms of the phasing coefficients. 
In an ideal scenario, one would like to treat all the eight phasing parameters as independent as suggested in Ref.~\cite{AIQS06a}. However, since all of them are functions of binary masses, there are strong correlations amongst them. This results in a near-singular Fisher information matrix whose inversion cannot be trusted. Hence it was possible to obtain reliably the errors on all the parameters only for a limited range of masses, within this approach~\footnote{{ Parameter estimation techniques like Bayesian inference can help to deal with such situations in the case of real data~\cite{NichVech98}.}}.

In the present work, we use the {\it singular value decomposition (SVD)} approach to circumvent this ill-conditioning of the Fisher matrix. Using SVD, we construct new parameters (corresponding to the dominant singular values of the Fisher matrix) which are combinations of the original phasing parameters, and obtain the errors associated with these new parameters. We find these errors 
to be very small which implies that one will be able to extract these new parameters with very good accuracy with ground based GW detectors such as aLIGO and ET. For example  we focus on the first three dominant new phase parameters. We show that these new phase parameters of GW can be estimated with relative accuracy of a few times $10^{-5}$ as emitted by a stellar mass BHs at 100 Mpc. Further,
for the Einstein Telescope, the accuracy for the stellar mass and intermediate BH binary inspirals at 100 Mpc would be $10^{-6}-10^{-5}$ which would enable one to very precisely estimate these parameters.

 This paper is organised in the following way. Sec.~\ref{PTPN} describes the parametrised tests of PN theory and discusses the waveform model we use.  A summary of the noise model and the parameter estimation scheme using Fisher matrix is presented in Sec.~\ref{Fisher}. Singular Value Decomposition of the Fisher matrix is explained in detail in Sec.~\ref{SVD}.
Our results are discussed in Sec.~\ref{Results}. Conclusions and future directions are elaborated in Sec.~\ref{Conclusion}.

\section{Parametrised Tests of PN theory}
\label{PTPN}

Fourier domain gravitational waveforms from non-spinning inspiralling compact binaries in circular orbit, obtained using stationary phase approximation, may be written as
\begin{equation}
\label{hf}
\tilde h(f) = {\cal A}\, f^{-7/6} e^{2 i \Psi(f)},
\end{equation}
with the amplitude ${\cal A}$  and phase $\Psi(f)$ given by\footnote{Throughout the paper we work in geometrical units with G=c=1}
\begin{eqnarray} 
{\cal A} & = & \frac{\cal C}{D_L\pi^{2/3}} \sqrt{\frac{5}{24}} {\cal M}^{5/6},\nonumber\\
\Psi(f) & = & 2\pi f t_c - \Phi_c +  \frac{\pi}{4} + \sum_{k=0}^7 [\psi_k +\psi_{kl}\,\ln f] f^{(k-5)/3}.
\label{eq:Fourier Phase}
\end{eqnarray} 
Here $t_c$ and $\Phi_c$ are the fiducial epoch of merger and the
phase of the signal at that epoch. Chirp mass ${\cal M}$ is related to the total mass $M$ by ${\cal M}={M\,\eta^{3/5}}$ where $\eta={m_1\,m_2\over(m_1+m_2)^2}$ ($m_1$ and $m_2$
are individual masses of the binary) is called the symmetric mass ratio which is $0.25$ for equal mass binaries. ${\cal C}$ which appears in the equation for amplitude of the waveform is a constant with $0\leq {\cal C} \leq 1$ which depends on the sky position of the source and the geometry of the detector and has a root-mean-square value of $2/5$ when averaged over all sky locations and source orientations. The coefficients in the PN 
expansion of the Fourier phase are given by~\cite{DIS01,DIS02,AISS05}:
\begin{equation}
\label{psik}
\psi_{k} = \frac{3}{256\,\eta}(\pi\, M)^{(k-5)/3}\alpha_{k} \,,\hspace{0.5in}
\psi_{kl} = \frac{3}{256\,\eta}(\pi\, M)^{(k-5)/3}\alpha_{kl}.
\end{equation}
where $\alpha_k$ and $\alpha_{kl}$ up to 3.5PN read as \cite{AISS05,MAIS10},
\begin{eqnarray}
\alpha_0 &=&1,\\ 
\alpha_1 &=&0,\\
\alpha_2 &=&\frac{3715}{756}+\frac{55}{9}\eta,\\
\alpha_3 &=&-16 \pi,\\
\alpha_4 &=&\frac{15293365}{508032}+\frac{27145}{504} \eta+\frac{3085}{72}
\eta^2; \\
\alpha_{5} &=&\pi\left(\frac{38645}{756}-\frac{65}{9}\eta\right)\,\left[1+\ln\left(6^{3/2}\pi M\right)\right],\\
\alpha_{5l} &=&\pi \left(\frac{38645}{756}-\frac{65}{9}\eta\right), \\
\alpha_6 &=&\frac{11583231236531}{4694215680}-\frac{640}{3}\pi
^2-\frac{6848}{21}C
+\left(-\frac{15737765635}{3048192}+ \frac{2255}{12}\pi ^2\right)\eta\nonumber\\
&&+\frac{76055}{1728}\eta^2-\frac{127825}{1296}\eta^3-\frac{6848}{63}\ln\left(64\,\pi M\right),\nonumber\\
\alpha_{6l}&=&-\frac{6848}{63},\\
\alpha_7 &=&\pi\left( \frac{77096675}{254016}+\frac{378515}{1512}
\eta -\frac{74045}{756}\eta ^2\right),\\
\alpha_{kl}&=&0,\, {\rm for }\;\; k=0,1,2,3,4,7 ,
\label{eq:phasingcoefficients2}
\end{eqnarray} 
where $C=0.577...$ is the Euler constant. It is evident from Eq.~(\ref{eq:Fourier Phase}) that different PN order terms occur with different frequency dependencies. For non-spinning binaries, each of the PN coefficients $\psi_k$ and $\psi_{kl}$ depends only on the component masses of the binary as given in Eq.~(\ref{psik}). Interestingly, from Eq.~(\ref{eq:Fourier Phase}), the non-logarithmic term corresponding to $k=5$ can be absorbed into a redefinition of the fiducial phase $\Phi_c$ as it has no frequency dependence. Hence there are eight parameters that characterise the phasing formula which are $\{\psi_0,\psi_2,\psi_3,\psi_4,\psi_{5l},\psi_{6l},\psi_6,\psi_7\}$.

The usual parameter estimation problem addresses the estimation of individual masses of the binary (or particular combinations of the component masses) using the 3.5PN accurate phasing formula. Two of the phasing coefficients are
sufficient to estimate the two component masses. The additional phasing coefficients improve this estimation~\cite{CF94,PW95,AISS05,ALISA06}.
 Alternatively, one can treat each one of the phasing coefficients as independent parameters in the parameter estimation problem. If this is possible, it will enable us to  test the PN structure of GR by independently measuring each PN coefficient and comparing with the predictions of PN approximation to GR~\cite{AIQS06a}. Though this idea works in principle, there will be large correlations amongst the different PN phasing coefficients because all of them are essentially functions of the two component masses of the binary.

To circumvent this problem, the proposal in~\cite{AIQS06b} uses only three out of eight parameters as independent in the parameter estimation problem. This is achieved by re-expressing  all the phasing coefficients, except the one to be used as test parameter,  in terms of two basic parameters (usually, the  two lowest order phasing coefficients). The third parameter used as test can be any of the higher order terms. Hence, authors proposed many such tests corresponding to different higher order phasing coefficients which are used as test parameter. Needless to say this test is not as robust as the original test where all of the phasing coefficients were independent, but this is one of the plausible ways to overcome the problem arising from correlations between various parameters.

This paper paves foundation for a new approach of working with the original proposal where all of the phase parameters are treated independently. Using SVD, one can get rid of the correlations among the PN phasing coefficients and obtain the most dominant new phase parameters as a linear combination of the original phase parameters. PN theory has unique predictions about the values of these new parameters and this can be compared with observations by directly measuring them from the data. Hence these parameters can be used to do tests of PN theory. 

\section{Detector noise model and parameter estimation using Fisher matrix}\label{Fisher}

We model the noise to be a Gaussian and stationary, random process. We are interested in the estimation of parameters of a signal buried in the noise. We use the Fisher Information matrix formalism~\cite{Rao45,Cramer46} to estimate the errors associated with the measurement of the signal parameters (see ~\cite{Finn92,FinnCh93} for some of the initial applications of Fisher matrix approach in the GW data analysis context. See a recent article about the caveats of using Fisher matrix as an error estimator~\cite{Vallisneri07}. A method to go beyond Fisher matrix is discussed in ~\cite{VZ10, Vallisneri11}.)

Let $\tilde{\theta}^{a}$ denote the `actual values' of the parameters 
and $\hat{\theta}^a \equiv \tilde{\theta}^{a}+\Delta\theta^{a}$  the best-fit parameters (or parameter estimate) in presence of noise. Then, in the limit of large SNRs, errors in the estimation of parameters ${\bf \Delta \theta} \equiv \{\Delta\theta^{a}\}$ obey a multivariate Gaussian probability distribution of the form
\begin{equation}
p({\bf \Delta\theta})={\cal N} \exp[{-\Gamma_{bc}\Delta\theta^{b}\Delta\theta^{c}}/2],
\label{eq:prob-dist}
\end{equation}
where ${\cal N}$ is the normalisation constant. The repeated indices are summed
over. The quantity $\Gamma_{ab}$ appearing in Eq.~(\ref{eq:prob-dist}) is the {\it
  Fisher information matrix} which is given by,
\begin{equation}
\Gamma_{ab}=(h_{a}\,|\,h_{b})
\label{Fishermatrix}
\end{equation}
where $h_{a}\equiv \partial h/\partial \theta^a$.
Here, $(\,|\,)$ denotes the noise weighted inner product as defined below.

Given any two functions
$a$ and $b$ their inner product is defined as:
\begin{equation}
(a\,|\,b)\equiv 4 \Re \int_{f_{min}}^{f_{max}}\,\frac{\tilde a^{*}(f)\,\tilde b(f)}{S_h(f)} df,\label{eq:inner product}
\end{equation}
where $S_h(f)$ is the one sided noise power spectral density (PSD) of the
detector. Hence the explicit expression for the Fisher matrix is given by,
\begin{equation}
\label{gamma}
\Gamma_{ab} = 4 \Re \int_{f_s}^{ f_{lso}}
\frac{\tilde{h}_{a}^*(f) \tilde{h}_{b}(f)}{S_h(f)}\; df.\label{eq:gamma-eqn}
\end{equation}
where $f_s$ is the lower cut-off frequency as given by the seismic noise limit and $f_{lso}$ is the upper cut-off frequency which is the frequency at the last stable orbit beyond which PN approximation breaks down. The frequency at the last stable orbit can be expressed as a function of the total mass $M$ of the binary as $f_{lso}={(6^{3/2}\pi\,M)}^{-1}$. $\Re(.)$ denotes the real part of the quantity in brackets.

The covariance matrix is the inverse of the Fisher matrix as given below
\begin{equation}
{\bf C} \equiv \{\langle \Delta \theta^a
\Delta \theta^b \rangle \} = {\bf \Gamma}^{-1},
\label{sigma_a}
\end{equation}
where $\langle \cdot \rangle$ denotes an average over the 
probability distribution function in Eq.~(\ref{eq:prob-dist}). 
The root-mean-square error $\sigma_a$ in the estimation of the
parameters $\theta^{a}$ is
\begin{equation}
\sigma_a =
\bigl\langle (\Delta \theta^a)^2 \bigr\rangle^{1/2}
= \sqrt{C^{aa}}\,.
\label{eq:sigma_a}
\end{equation}
Note that in the above expression, index $a$ is {\it not} summed over.

For the noise power spectral densities  of aLIGO and ET we have used  the analytic fit
given in Eq.~(2.1) of \cite{MAIS10} and Eq.~(2.2) and (2.3) of \cite{MAIS10}, respectively. The lower limit of the frequency is assumed to
be given by the seismic cut off which is taken to be 20Hz for aLIGO and 1 Hz for ET.
\section{Singular value decomposition and Fisher information matrix}\label{SVD}
Singular Value Decomposition is a well-known technique in signal processing to transform a set of  correlated variables into
a set of uncorrelated variables in terms of which certain features of the data are more evident. This technique is used both for
detection purposes (see for example ~\cite{CHK11}) and in studies of parameter estimation~\cite{SSCQG03} in the case of GW data analysis.
There are techniques similar to SVD which have also been used in the case of detection~\cite{FieldRedBasis11,FrankRedBasis12}.

In this section we recast the parameter estimation problem with the help of singular value decomposition of the Fisher matrix.
 We show that the truncated SVD 
eliminates the near-singular feature of the Fisher information discussed
in Sec. \ref{PTPN}. To demonstrate, we opt to work in the discrete frequency domain.
However, the final result does not depend on this choice.  

Let the  ${\bf \theta} \equiv \{\theta^i \}$ denotes the $m$ dimensional vector corresponding to the PN phase  parameters as defined in Eq.~(\ref{psik}). We divide the frequency band $(f_s,f_{lso})$ into N discrete frequency points $f_k$ with bin-size $\Delta f$ such that $f_0=f_s$ and 
$f_{N-1}=f_{lso}$. In the discrete domain, Eq. (\ref{gamma}) is recast as
\begin{equation}
\Gamma_{ij} = 4 \Re \left[ \sum_{k=0}^{N-1} \frac{{\tilde{h}}_{,i}(f_k)}{\sqrt{S_h(f_k)}}
\frac{{\tilde{h}}^*_{,j}(f_k)}{\sqrt{S_h(f_k)}} \right]~\Delta f \,.
\end{equation}
where at any frequency bin $f_k$, ${\tilde{h}}_{,i}(f_k) = \partial{\tilde{h}}(f_k)/\partial \theta^i$ denotes the $i^{\rm th}$ component of the tangent
vector of the GW signal with respect to the phase parameters.

We define ${\bf H}=\{H_{ik} \equiv \frac{{\tilde{h}}_{,i}(f_k) \sqrt{2~\Delta f}}{\sqrt{S(f_k)}} \}$ to be  a matrix of size $m \times N$ where the
row index $i$ corresponds to the phase parameter space and the column index  $k$ is for the frequency index.  The matrix ${\bf H}$ physically signifies the variation of the signal in the $m$ dimensional parameter space ${\bf \theta}$ when evaluated at different frequency locations. It further acts as a building block for Fisher matrix as shown below.

The Fisher matrix re-expressed in terms of ${\bf H}$ takes the form 

\begin{equation}
\label{gamma2}
\Gamma_{ij} = \sum_{k=0}^{N-1} [ H_{ik} H^*_{jk} + H^*_{ik} H_{jk} ] \,.
\end{equation}

 The main contribution to the variation of the signal in the parameter
space comes from the variation of the phase. Hence, owing to the form of the signal 
expressed in Eq. (\ref{hf}), we note that $H_{ik} H^*_{jk}$ is a real quantity in the phase parameter space ${\bf \theta}$. In terms of ${\bf H}$, the $m \times m$ Fisher matrix compactifies to 
\begin{equation}
\label{gamma3}
{\bf \Gamma} = 2~{\bf H H^\dag} \,.
\end{equation}

Henceforth, we focus on ${\bf H}$. The singular value decomposition (SVD) is a powerful tool to identify the subspace of $m\times N$ space where ${\bf H}$ resides or in other words, the directions along which the signal varies dominantly in the parameter space at a given frequency location.

The method of SVD provides a factorisation of any rectangular matrix into a square matrix, a rectangular diagonal matrix and another square matrix as given below
\begin{equation}
\label{svdH}
{\bf H = U_H \Omega_H V_H^\dag} \,,
\end{equation}

\noindent
where ${\bf U_H}$ and ${\bf V_H}$ are the left and right unitary square matrices of $m$ and $N$ dimensions respectively and the rectangular ${m \times N}$ diagonal matrix  ${\bf \Omega_H}$ arranges the singular values of ${\bf H}$
in the descending order. The columns of ${\bf U_H}$ and ${\bf V_H}$ form an orthonormal basis set
 in $m$ and $N$ dimensional space respectively {\it i.e.} ${\bf U^\dag U = V^\dag V = I }$.
Further, they act as (orthogonal) eigen-vectors of $({\bf H H^\dag})_{m\times m}$
and $({\bf H^\dag H})_{N \times N}$ respectively. Substituting Eq. (\ref{svdH}) in Eq. (\ref{gamma3}), Eq. (\ref{gamma3}) emerges as a SVD of ${\bf \Gamma}$ as

\begin{equation}
\label{Gamma4}
{\bf \Gamma = U_H \Sigma U_H^\dag} \hspace{0.5in} \Rightarrow \hspace{0.5 in}  {\bf \Gamma U_H = \Sigma U_H} \,,
\end{equation}
where ${\bf \Sigma = 2\, \Omega_H \Omega_H^\dag}$  with the diagonal entries as $\Sigma_{kk} = 2 (\Omega_H)_{kk}^2$.

We note that {\it the  {\bf $U_H$} provides the SVD for the Fisher matrix with its columns act as eigen-vectors of the ${\bf \Gamma}$ }. In other words, they provide the principal components of ${\bf \Gamma}$. Hence we have shown an alternative way to obtain the principal components of Fisher matrix using the SVD of ${\bf H}$. We can compare and contrast our method with the principal component analysis (PCA) used, for example, in Ref.~\cite{SSCQG03}.

In Ref.~\cite{SSCQG03}, the authors focuses on the PCA of the covariance matrix in the context of parameter estimation of spin modulated chirp signals in the LISA band \footnote{Laser Interferometer Space Antenna (LISA) is a proposed space-based GW detector sensitive to the low frequency band between $\sim 10^{-4}-1$ Hz}. Since the covariance matrix is simply the inverse of the Fisher matrix, both the methods would be equivalent for a non-singular ${\bf \Gamma}$.

\subsection{Treating  the ill-conditioning of the Fisher Matrix}

To recall, in order to test PN theory, the ideal scenario would be to measure
all the PN phase coefficients independently which was not possible due to large correlations among the parameters. 
The main hurdle to address the full problem was that the Fisher matrix when
expressed in the 8-dimensional parameter space  ${\bf \theta} \equiv \{\psi_0,\psi_2,\psi_3,\psi_4,\psi_{5l},\psi_{6l},\psi_6,\psi_7 \}$,  is ill-conditioned.
 An ill-conditioned Fisher Matrix implies that ${\bf \Gamma}$ cannot be inverted reliably. For cases where ${\bf \Gamma}$ is near-singular, SVD approach  provides an elegant solution to deal with the singularity and conditions the Fisher matrix.
Rest of the section, we shall focus on this.

 Consider the parameter space to be  ${\bf \theta} \equiv \{\psi_0,\psi_2,\psi_3,\psi_4,\psi_{5l},\psi_{6l},\psi_{6},\psi_7 \}$ {\it i.e.} $m=8$. We compute
the SVD of ${\bf H}$ in this parameter space. In case of the near-singular
matrix, some of the diagonal entries of ${\bf \Sigma}$ are very close to zero. As a result, the errors in those parameters are very large or unreliable. Furthermore, the effect of singular nature of ${\bf \Gamma}$ introduces large errors in other parameters as well.
 Thus, to obtain the reliable errors,  the first step is to condition the Fisher matrix  or treat its singular nature. There are various approaches
to address this problem. We follow the {\it Truncated SVD} approach in this paper \cite{SVD}.

The truncated SVD approach is elaborated in our context as below. We first
perform the SVD of ${\bf H}$. We then examine the singular values given by the diagonal entries of ${\bf \Omega_H}$ i.e. $(\Omega_H)_{kk}$. The total number of non-zero diagonal entries gives the rank of ${\bf H}$. We define the ratio of the singular values $(\Omega_H)_{kk}/(\Omega_H)_{11}$ and consider only those which are greater than or
equal to a small number $\epsilon$. Let the total number which satisfy this criterion be $r$. Then we truncate $\Omega_H$ to the dimensions
equal to $r$. The new truncated singular matrix ${\bf \Omega_H^t}$ is $r \times r$ with diagonal entries.
Consequently, the truncated version of left and right singular
matrices  ${\bf U_H^t}$ and ${\bf V_H}$ is a subset of ${\bf U_H}$ and 
${\bf V_H}$ respectively which corresponds to the first $r$ dominant singular values. The truncated SVD takes the form
\begin{equation}
{\bf H = U_H^t \Omega_H^t V_H^{t\dag}}
\end{equation}
where the ${\bf U_H^t}$ is $8 \times r$, ${\bf \Omega^t}$ is $r \times r$ and
${\bf V_H^{t\dag}}$ is $r \times N$. Thus, in truncated SVD approach $8 \times N$ dimensional ${\bf H}$ 
is reconstructed by eliminating all non-dominant/near-zero singular values which otherwise do not contribute to
${\bf H}$. This explicit elimination of zeros, brings us into a subspace in which ${\bf H}$ resides. Hence, 
by construction, the truncated SVD given in Eq.~(\ref{Gamma4}) provides conditioned invertible Fisher matrix
\begin{equation}
\label{Gamma3}
{\bf \Gamma = U^t_H \Sigma^t U_H^{t\dag}}\hspace{0.5in} {\Rightarrow}\hspace{0.5in} {\bf  U_H^{t\dag} \Gamma U_H^t} = {\bf \Sigma^t}
\end{equation}
where ${\bf \Sigma^t = 2 \Omega_H^t \Omega_H^{t\dag}}$ and ${\bf U_H^t}$ give 
Eigen-vectors corresponding to the dominant Eigenvalues. Thus the truncated SVD conditions the  ${\bf \Gamma}$. In fact, ${\bf U_H^t}$ provides the unitary transformation Eq.~(\ref{Gamma3}) which diagonalises ${\bf \Gamma}$ with the diagonal entries ${\bf \Sigma}_{kk} = 2 (\Omega_H^t)_{kk}^2$ for $k=1,\ldots,r$.

\begin{figure}[t]
\begin{center}
\includegraphics[scale=0.4]{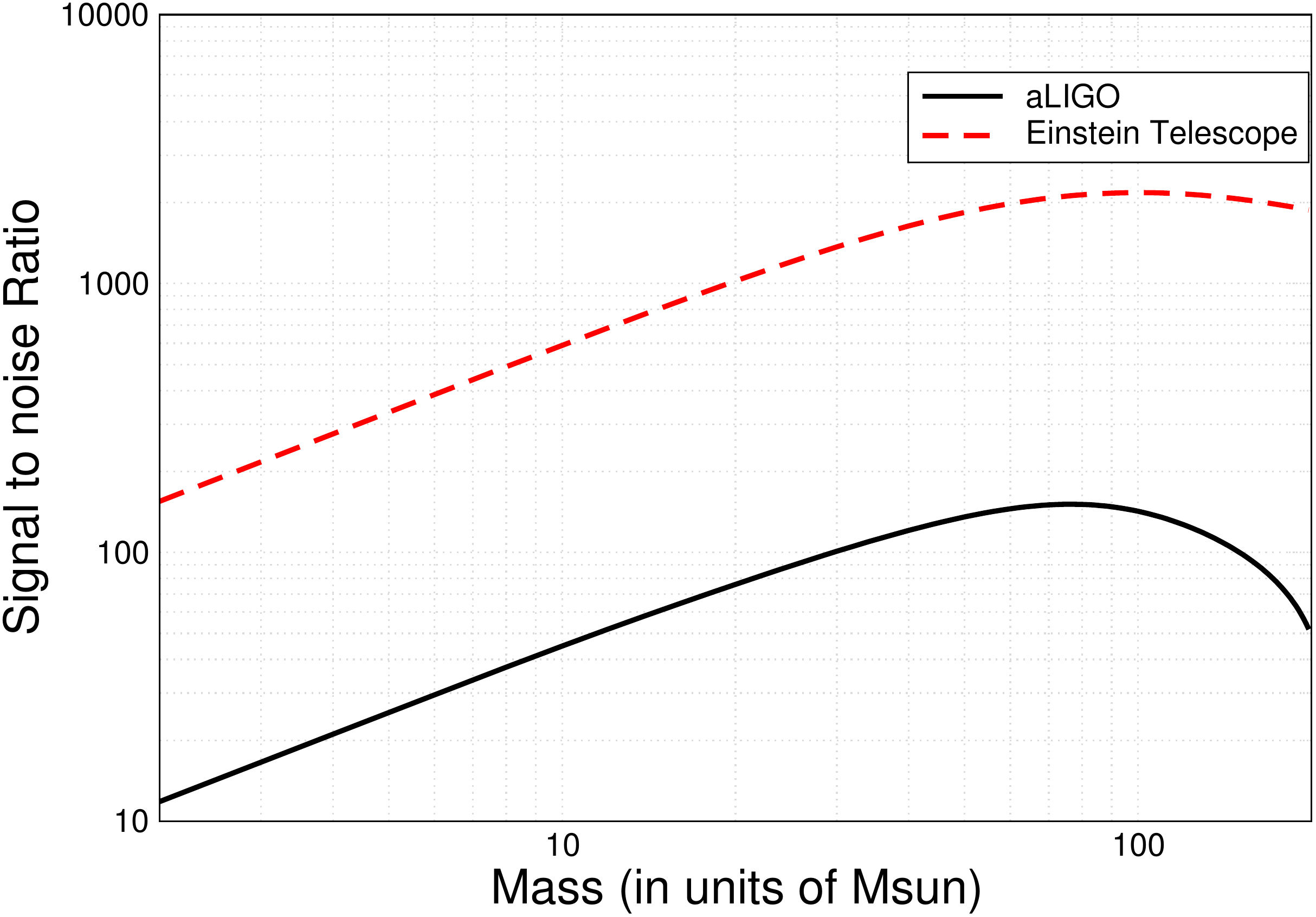}
\end{center}
\caption{Optimal signal to noise ratio (SNR) as function of total mass. The figure is for equal mass binaries. Sources are assumed to be at a fixed distance of 100 Mpc irrespective of the mass. A seismic cut-off frequency of 20Hz is assumed for aLIGO and 1 Hz for ET. Einstein Telescope has about ten times the signal to noise ratio compared to aLIGO for the same total mass. Pattern averaged waveforms introduced in Sec.~\ref{PTPN} is used. (One should bear in mind that, the actual SNRs of aLIGO sources will go inversely as distance and the distance of 100 Mpc we have assumed is more for analytical convenience and not based on any astrophysical model. Hence the actual SNRs of aLIGO may be much smaller than what is shown in the figure.)}\label{fig:SNR}
\end{figure}
 
Comparing this approach with that of Ref.~\cite{SSCQG03}, one observes that the latter is based on the PCA of the covariance matrix which is the inverse of Fisher matrix. This method cannot be applied if the inversion of the Fisher matrix itself is unreliable, as is in our case. In other words, PCA is not aimed at removing the near singular nature of the Fisher matrix whereas our procedure is specifically devised for that.
\subsection{New Phase Parameters and the relative errors}
Let $\Delta \theta^{\alpha}$ be the error in the phase parameters $\theta^{\alpha}$ which obey the multi-dimensional Gaussian distribution following Eq. (\ref{eq:prob-dist})
as given below
\begin{equation}
p(\Delta {\bf \theta})={\cal N} \exp[{-\Gamma_{\alpha \beta}\Delta \theta^{\alpha}\Delta \theta^{\beta}}/2],
\label{eq:prob-dist2}
\end{equation}
The repeated indices are summed over and $\alpha,\beta = 1,\ldots 8$.

Consider the scalar $\Gamma_{\alpha \beta }\Delta \theta^{\alpha}\Delta \theta^{\beta}$ in Eq. (\ref{eq:prob-dist2}).
The truncated SVD approach should keep this scalar invariant as we are not removing
any information but removing the zeros in the problem. We use this fact to obtain the new
phase coordinates. 
\begin{eqnarray}
\label{eq:prob_dist3}
\Delta {\bf \theta}^T {\bf \Gamma} \Delta \bf{\theta} &=& \Delta {\bf \theta}^T \, {\bf U^t_H \Sigma^t U_H^{t\dag}} \, \Delta {\bf \theta} \nonumber \\
&\equiv& \Delta {\bf \theta'}^T \, {\bf \Sigma^t} \, \Delta {\bf \theta'} 
\end{eqnarray}
where superscript $T$ shows the transpose of the matrix.
The new phase coordinates ${\bf \theta'}$ are obtained by linearly combining the
PN phase parameters ${\bf \theta}$ as
 \begin{equation}
{\bf \theta' = U^{t \dag}_H \theta} \,.
\end{equation}
where  ${\bf \theta'} \equiv \{\psi'_1,\psi'_2,\psi'_3,\ldots,\psi'_r\}$ is a $r<8$ dimensional phase vector. 
The Fisher matrix and hence the covariance matrix is diagonal in this new phase coordinates. Thus, ${\bf \theta'}$
phase coordinates are statistically independent.

\begin{figure}[t]
\begin{center}
\includegraphics[scale=0.30]{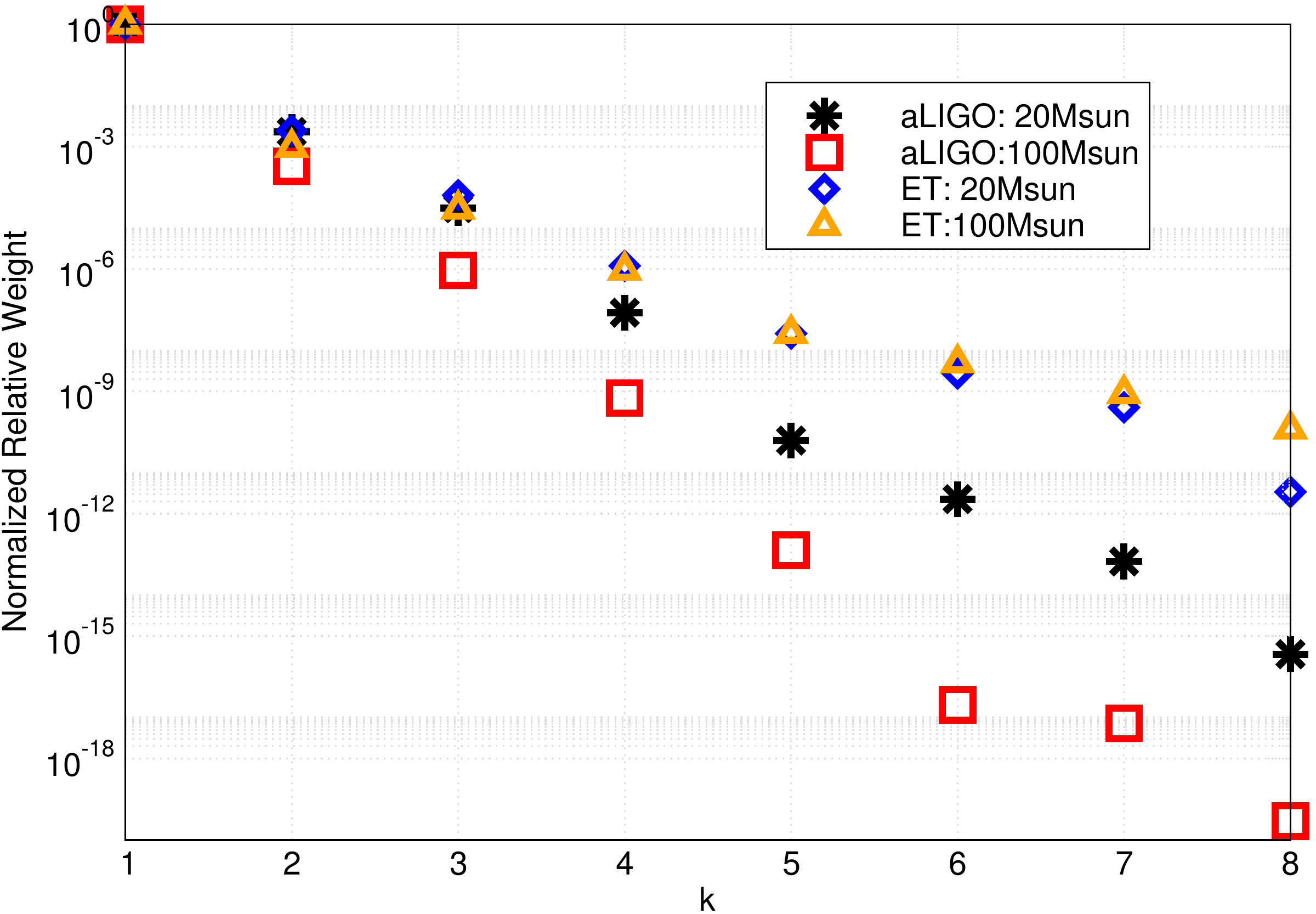}
\includegraphics[scale=0.30]{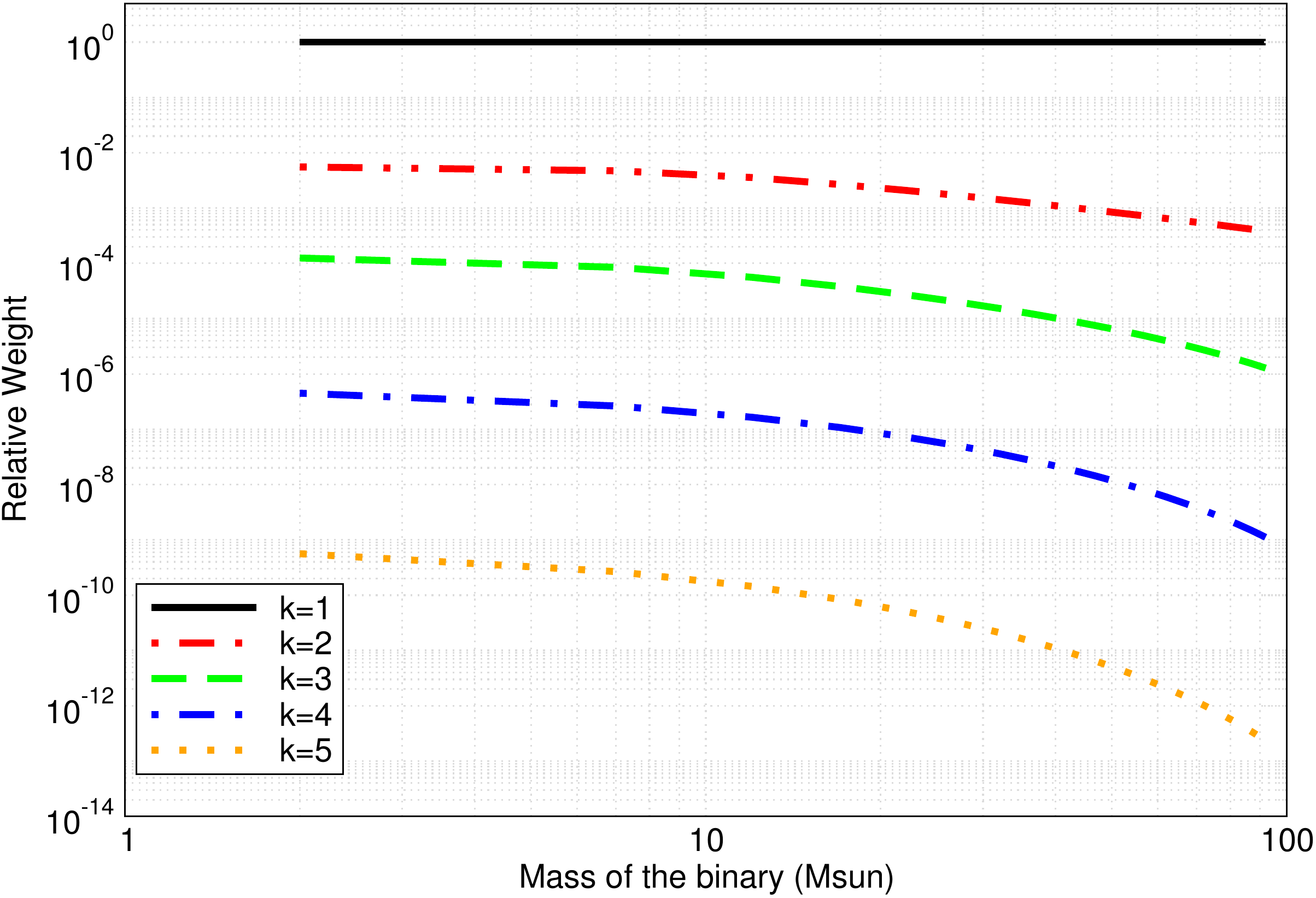}
\end{center}
\caption{{\bf Left panel:} Ratio of  singular values of the Fisher matrix w.r.t the dominant one ($\frac{\Sigma_{kk}}{\Sigma_{11}}$) as a function of $k$, where $k$ denotes the different singular values and runs from 1 to 8. {\bf Right Panel:} For the case of aLIGO detector, variation of relative weights as a function of total mass for various $k$ values. Both the panels are for equal mass systems ($\eta = 0.25$).}\label{RelWeights}
\end{figure}

Let the errors in parameter estimation be $\Delta {\bf \theta'}$. Owing to the statistical
independence, the multi-dimensional Gaussian distribution of $r$-independent random variables
$\Delta {\bf \theta'}^\alpha$ is the product of the $k$ Gaussian probability distributions with zero mean and
variance $\sigma_\alpha = (\Sigma^t_{\alpha \alpha})^{-1/2}$ as shown below
\begin{equation}
p(\Delta {\bf \theta'})={\cal N'}\, \Pi_\alpha \exp[-{(\Delta \theta'^{\alpha})}^2/(2 \sigma_\alpha^2)],
\label{eq:prob-dist4}
\end{equation}
where $\alpha = 1,\ldots r$ and ${\cal N'}$ is a normalisation constant.

\subsection{Phasing formula in terms of the new phase parameters}
The PN phase given in Eq.~(\ref{psik}) can be expressed in terms of ${\bf \theta}$ as
\begin{equation}
\Psi(f) =  2\pi f t_c -\Phi_c + \frac{\pi}{4} + \sum_{j=1}^8 P_j \theta_j.
\end{equation}
where $\sum_{j=1}^8 P_j \theta_j \equiv \sum_{k=0}^7 [\psi_k +\psi_{kl}\,\ln f] f^{(k-5)/3}$ such that 
\begin{equation}
{\bf P} \equiv \{ f^{-5/3}, f^{-1}, f^{-2/3}, f^{-1/3}, \ln f, f^{1/3}\,\ln f, f^{1/3}, f^{2/3} \}
\end{equation}
is a row vector as function of frequency.
Using the unitary transformation, ${\bf \theta' = U^{{\dag}}_H \theta}$ and
using ${\bf U U^{{\dag}} = I}$, one can write the sum $\sum_{j=1}^8 P_j \theta_j$ as
\begin{equation}
{\bf P \theta = P U U^{{\dag}} \theta = P' \theta'} \,.
\end{equation}
Thus, in terms of the new phase parameters the PN phase becomes
\begin{equation}
\Psi(f) =  2\pi f t_c - \Phi_c +  \frac{\pi}{4} + \sum_{k=1}^8 P'_k \theta'_k.
\end{equation}

In the next section, we give comparison of how various $\psi_k$s behave as function of the total mass (for equal mass systems) in the original and new representations; see e.g. Fig.~\ref{OldNewPsik}.

\section{Results and Discussions}\label{Results}
\begin{figure}[t]
\begin{center}
\includegraphics[scale=0.30]{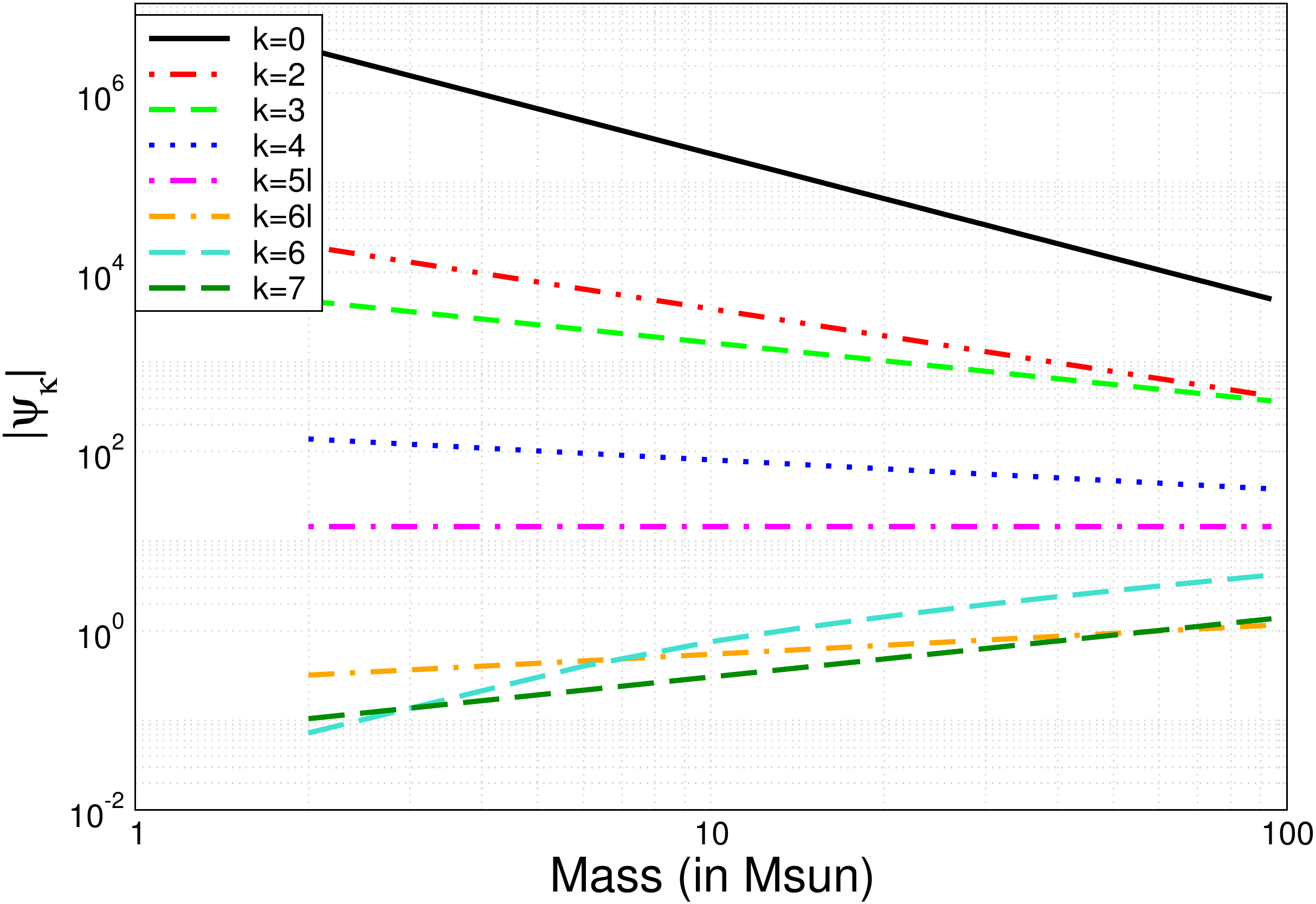}
\includegraphics[scale=0.30]{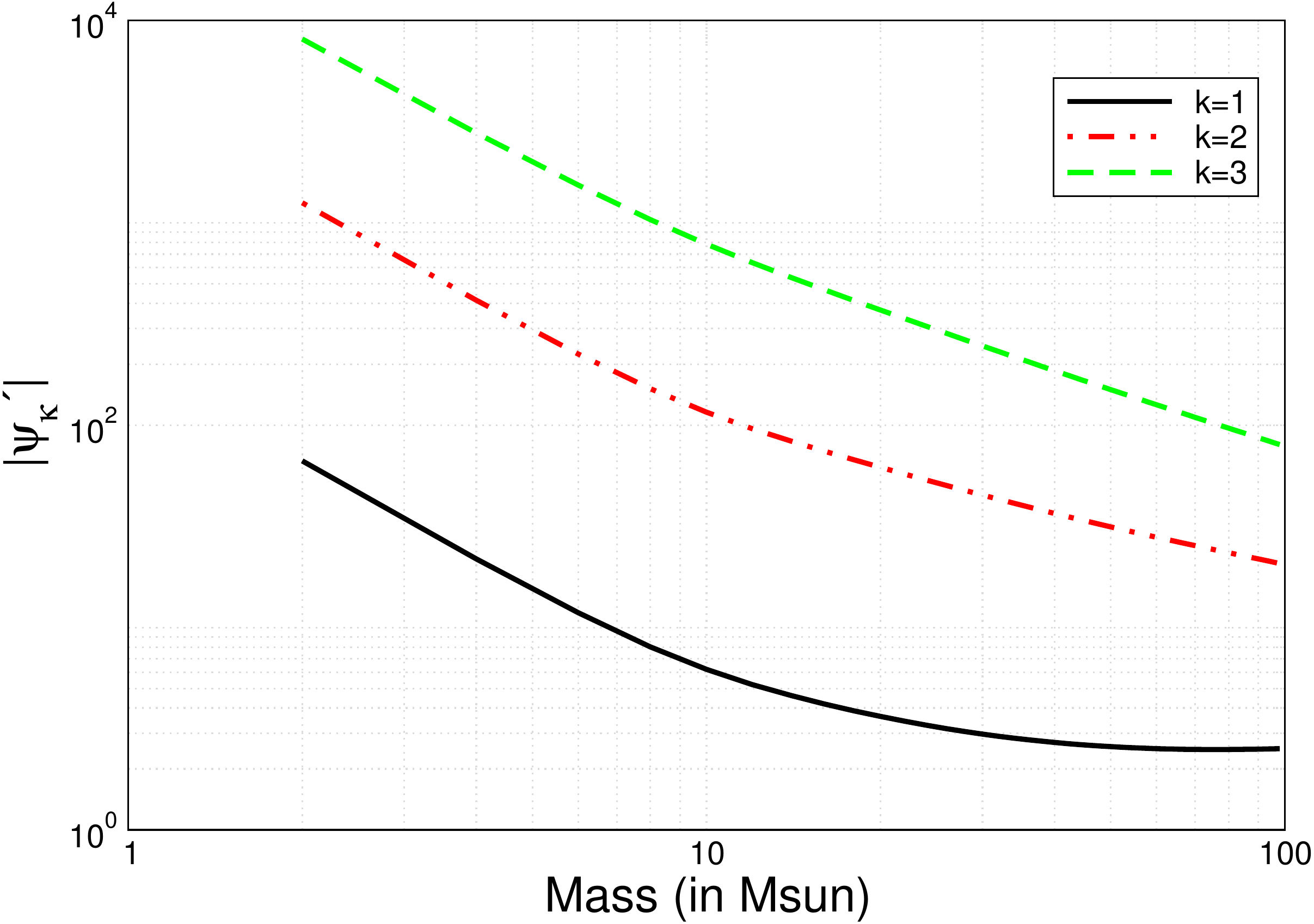}
\end{center}
\caption{$|\psi_k|$ (left panel) and $|\psi_k'|$ (right panel) as functions of the total mass of the binary for the equal mass case.\label{OldNewPsik}}
\end{figure}

 In this section, we demonstrate the outcome of using truncated SVD approach to the parameter estimation problem we were interested in and present
the errors on the new phase parameters.
Before we start discussing our results it is worth recalling the typical SNRs expected from the aLIGO and ET detections as a function of the total mass. This is presented in Fig.~\ref{fig:SNR}. Roughly speaking, ET events will have a SNR which is an order of magnitude higher than that of aLIGO; thanks to the smaller seismic cut-off and improved sensitivity. In the figure, the distance to the source is assumed to be 100 Mpc and we have considered binaries with equal masses.

We now proceed and compute the Fisher matrix for a particular
binary system with a fixed mass following the prescription outlined in Sec.~\ref{Fisher}. This amounts to computing the derivatives w.r.t the various parameters (the phasing coefficients) and performing the integration in Eq.~(\ref{eq:gamma-eqn}). We then obtain the matrix ${\bf H}$ and carry out the SVD using the in-built Matlab function. We observe that if we set  $\epsilon \sim 10^{-3}$, the first 3 singular values are non-zero with the rest of the diagonal entries decreasing rapidly for all total masses considered.
 We retain only the first three ($k=1,2,3$) leading singular values of ${\bf \Sigma}$ and truncate ${\bf \Sigma}$ to a 3 $\times$ 3 matrix. 
As was shown earlier, the above procedure is equivalent to first diagonalise the Fisher matrix and then rank
the diagonal entries in the descending order and retain only the first three 
eigenvalues, ignoring the rest. Both the procedures are numerically tested
to be equivalent.  Values of the diagonal entries (normalised to 1) are plotted as a function of the  index of the singular values in Fig.~\ref{RelWeights} for two different total masses and for aLIGO and ET configurations. From the figure it is obvious that eigenvalues corresponding to $k\geq 4$ are much smaller than the eigenvalues of the first parameter. This means the contribution of the parameters with $k\geq4$ is negligible in comparison with the first three parameters.
Thus we omit all eigenvalues higher than the first three. By diagonalising the Fisher matrix, we obtain a set of new parameters as linear combinations of the old parameters which are the new phasing coefficients.

In Fig.~\ref{OldNewPsik}, we plot the old and the new $|\psi_k|$ with the total mass of the binary. For the new phasing coefficients, we have shown the curves only for the dominant three given by the truncated SVD. The original phasing coefficients have essentially power law dependencies on the total mass and hence are linear in the log-log scale. The only exception to this is $\psi_6$ which has a term proportional to $M^{1/3}\ln M$ and hence not linear. The new parameters which are linear combinations of the old ones are not linear in the log-log scale. The other interesting thing to note is that the values of the old $|\psi_k|$ decrease with increasing $k$ whereas the new $|\psi_k|$ values increase with  $k$.

\begin{figure}[t]
\includegraphics[scale=0.3]{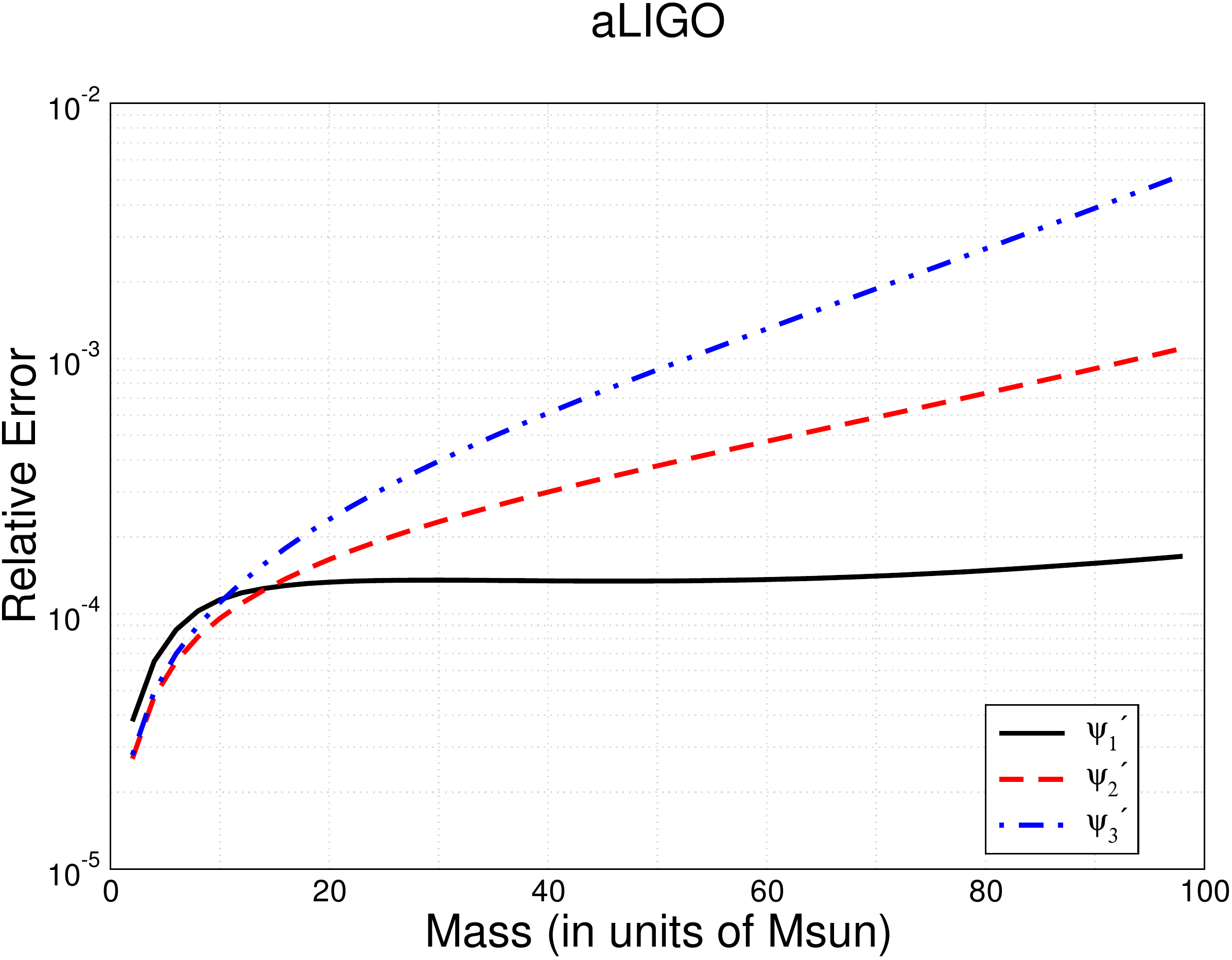}
\includegraphics[scale=0.3]{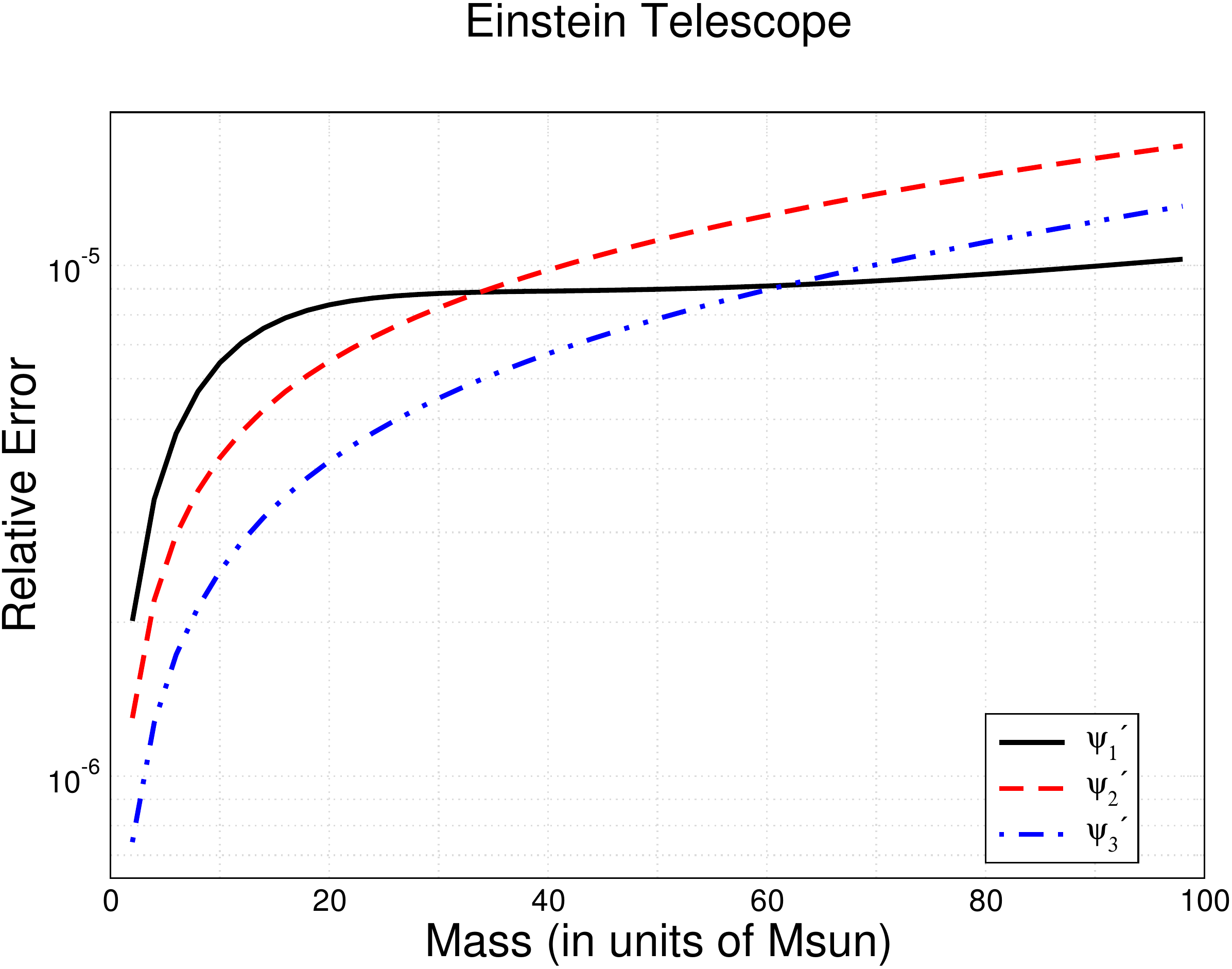}
\caption{Errors in the three new parameters (which are linear combinations of the old  parameters) obtained from SVD in the aLIGO and ET bands as functions of the total mass of the binary for equal mass systems $\eta = 0.25$. Distance to the source is assumed to be 100 Mpc irrespective of the total mass. Einstein Telescope errors are better by a factor of $\sim10$ due to the higher SNR ET sources have compared to aLIGO.}\label{fig:errors}
\end{figure}

Inverse of the diagonalised $3\times3$ Fisher matrix ${\bf \Sigma^t}$ directly gives the  covariance matrix corresponding to the new set of phase parameters ${\bf P'}$.  The square root of the diagonal entries of the covariance matrix yields the $1-\sigma$ error bars associated with the new set of three parameters.
\begin{figure}[t]
\includegraphics[width=0.5\textwidth]{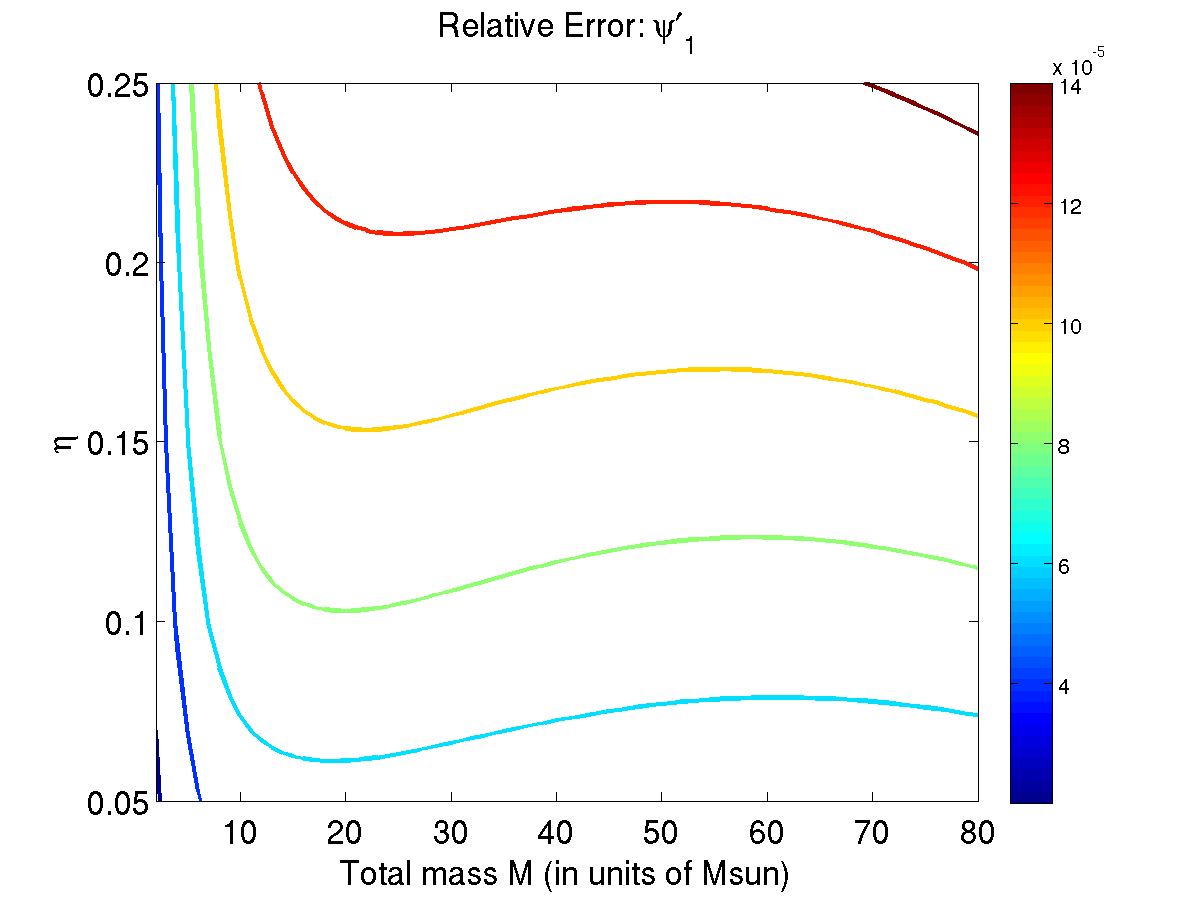}
\includegraphics[width=0.5\textwidth]{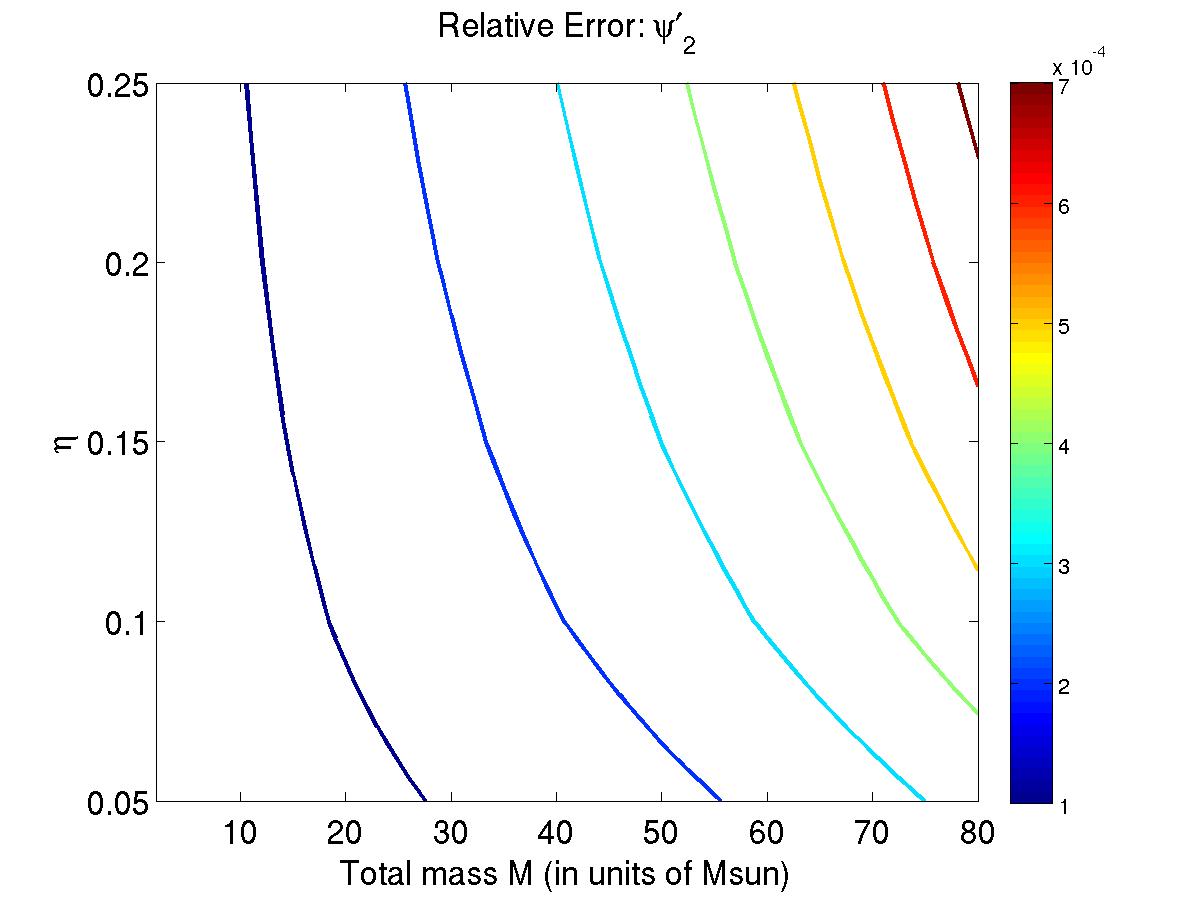}
\hspace*{1.5in}{~} \includegraphics[width=0.5\textwidth]{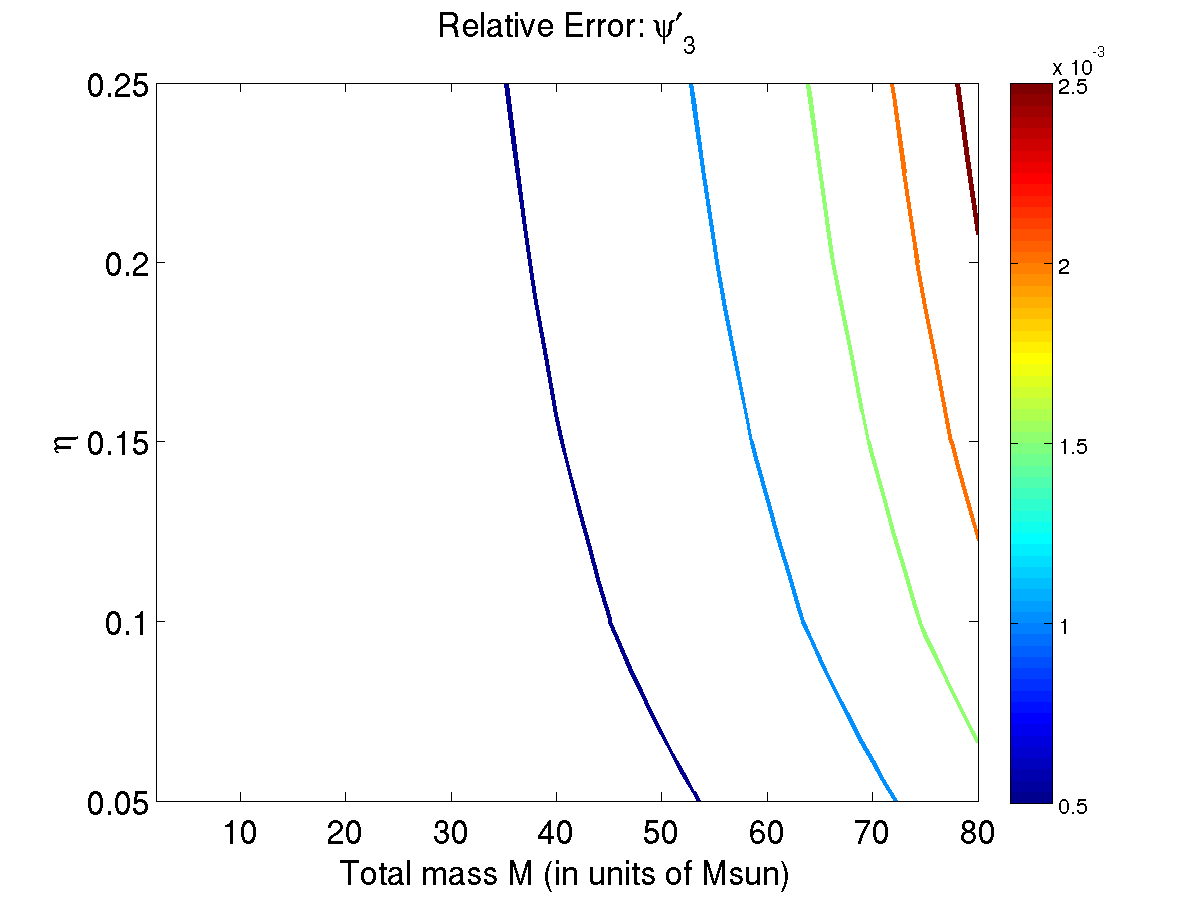}
\caption{Contour plots of Relative Errors in the three new parameters (which are linear combinations of the old  parameters) obtained from SVD in the aLIGO band. Distance to the source is assumed to be 100 Mpc irrespective of the total mass.
The relative errors for unequal mass cases are better than the equal mass case for fixed total mass.}\label{fig:errors_eta}
\end{figure}

In Fig.~\ref{fig:errors} we show the relative errors associated with the measurement of the three new parameters, 
as a function of mass for aLIGO and ET configurations. For masses $< 50M_{\odot}$, the relative errors in the new parameters (termed as $\psi_1', \psi_2', \psi_3'$) are $\leq 10^{-3}$ for aLIGO and $\leq 10^{-5}$ for ET. We note that removal of singularity improves the relative errors immensely. This is the primary result of this paper. 
This new set of parameters, which can be measured
more accurately, should be more useful for devising parametrised tests of PN theory.

At this stage it is worth comparing the relative errors we have obtained for the new phasing parameters with that of Ref~\cite{AIQS06b}. In Ref.~\cite{AIQS06b}, the authors used a reduced parameter set containing two basic parameters (which are the Newtonian and 1PN phasing coefficients) and a third test parameter which can be any one of the higher order phasing coefficient. They estimated the errors associated with the measurement of these three parameters. The best case scenario for the third generation EGO configuration (somewhat similar to the ET configuration we have considered here) has relative errors which ranges between $10^{-2}-1$ for most of the coefficients, the only exemption being that of $\psi_3$ which has a relative error of $10^{-3}$ (see left panel of Fig.~2 of Ref.~\cite{AIQS06b}). Comparing these results with right panel of Fig.~3 of this paper, one sees that all the three new phasing coefficients can be estimated with  relative accuracies $\sim 10^{-5}$ which is roughly 2-3 orders of magnitude better.
 This is solely due to the usage of the new set of parameters obtained using SVD.

It should be borne in mind, while looking at Fig.~\ref{fig:errors}, that what we have shown are only the {\it statistical errors} arising from the noise.
There can be systematic errors due to the fact that PN waveforms are only approximate. There have been proposals in the literature to quantify the
{\it systematic biases} due to inaccurate waveform modelling~\cite{CutlerVallisneri07}. We have not considered this aspect in the present work.

In Fig.~(\ref{fig:errors_eta}), we plot the contours of relative errors for a general case of unequal mass binaries in the $M-\eta$ plane. The colours of the contours show the values of the relative errors. From the  contours, it is clear that the approach works well for the entire range of $\eta$ values.
Throughout the parameter space, SVD approach can provide the three dominant parameters  (for $\epsilon \sim 10^{-3}$) which can be very well estimated using matched filtering.
To understand the relative error trends, we focus the reader's attention that
there are two competing factors that affect the relative errors; namely the SNR and the number of cycles spent in the detector band. For inspiral phase, SNR $\propto M^{5/6} \eta^{1/2}$. The number of cycles $N_{cycles} \propto M^{-5/3} \eta^{-1}$. 

\begin{figure}[t]
\includegraphics[width=0.5\textwidth]{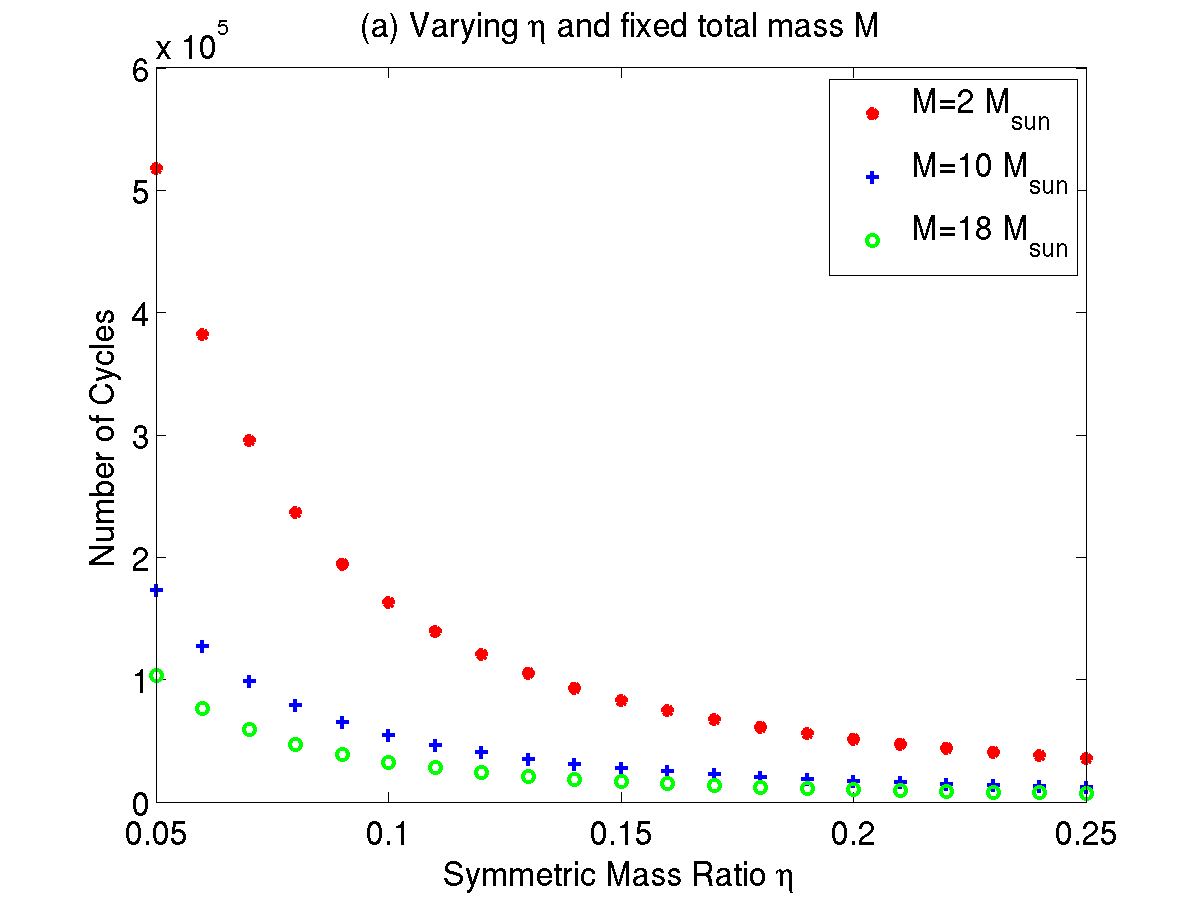}
\includegraphics[width=0.5\textwidth]{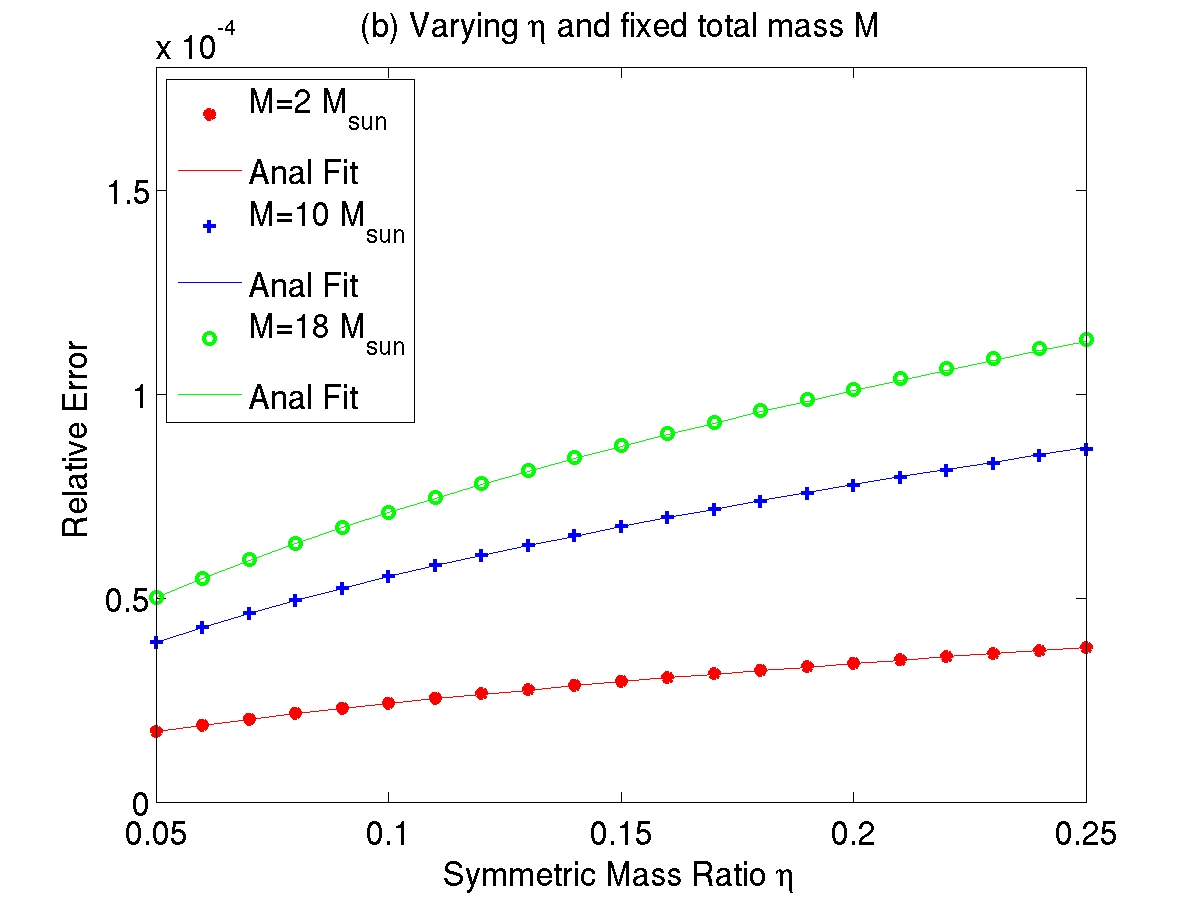}
\caption{Fixed total mass (2 $M_\odot$, 10 $M_\odot$, 18 $M_\odot$) and varying $\eta$ for aLIGO: (a) Number of cycles as a function of $\eta$. (b) Relative error of $\psi'_1$ as a function of $\eta$. The analytical fit for M=2 $M_\odot$, error = $7.5*10^{-5} \eta^{0.4925}$, for M=10 $M_\odot$, error = $17.25*10^{-5} \eta^{0.4947}$ and M=18 $M_\odot$, error = $22.8*10^{-5} \eta^{0.5066}$.\\
}\label{fig:errors_eta2}\end{figure}

For a fixed $\eta$ case, parameters of the binaries with low masses spend many more cycles in the detector band compared to the massive system. Though the SNR increases with the total mass, the effect of these two factors together makes the
relative error increase with total mass as $M^{5/6}$. On the contrary, for a 
fixed total mass and varying $\eta$ case, systems with smaller $\eta$ (more asymmetric) spend more time in the band (see left panel of Fig.~\ref{fig:errors_eta2}).
Here, the number of cycles are plotted as a function of $\eta$ which varies as $\eta^{-1}$ \cite{AISS05}. The SNR varies as $\eta^{1/2}$. The net product gives the relative error variation as $\eta^{1/2}$ which is inverse of the square root of the
number of cycles. In the right panel of  Fig.~(\ref{fig:errors_eta2}), we plot the relative error of $\psi'_1$ as a function of symmetric mass ratio $\eta$ for 
fixed total mass M for aLIGO noise. Analytical fit of the relative error as a function of $\eta$ shows that the relative error indeed varies as $\sim \eta^{0.5}$.     

Though we have considered both aLIGO and ET on equal footing for our calculations, we wish to emphasise that aLIGO is the upgraded version of
initial LIGO detector and is fully funded and instrumentation for which have already started. ET on the other hand is a proposed third generation
interferometer, still in the phase of design studies.

\section{Conclusions and future directions}\label{Conclusion}
We revisited the parameter estimation problem where each one of the PN phasing coefficients are treated as independent parameters and used to devise a self-consistency test of PN approximation to GR. We showed that using Singular Value Decomposition of the Fisher matrix, it is possible to deduce the three new combinations of the phasing coefficients which are best estimated using matched filtering.
The Fisher matrix and hence the covariance matrix are diagonal in this
new phase parameters which in turn are linear combinations of 
the original PN phasing coefficients. 
Pertaining to the removal of the singular nature, the new phase parameters can be
estimated very accurately with very small relative errors $\sim 10^{-3}-10^{-6}$ for aLIGO and ET configurations. This covers the BNS and NS-stellar mass BH cases for aLIGO and BNS, NS-BH and IMBH binaries for ET. We studied how the errors in the new parameters vary as a function of the total mass  of the binary as well as the symmetric mass ratio $\eta$.

However, here we have not discussed in detail how to rephrase the parametrised tests of PN theory in terms of these new parameters. This requires devising a consistency test using this new set of phasing parameters. One approach  could be the traditional approach to
follow the procedure outlined in Refs.~\cite{AIQS06a,AIQS06b,MAIS10} where
the consistency test(s) where carried out in the mass plane of the binary. One may carry out similar tests in the $m_1-m_2$ plane in terms of the three dominant parameters obtained using SVD.
Another approach could be  to fold in the new parameters in the
Bayesian framework following the the work of ~\cite{LiEtal2011} in which case the formulation and the implementation of the test is entirely different from the first one. Hence, our work should be seen as a first step towards the final goal of devising tests of PN theory where we have parameters which can be estimated much more accurately than those in the original proposal. These issues will be addressed in detail in future work.  We foresee that since these new phase parameters pertains to the principle components of the Fisher matrix which in turn is related to the maximal variation of the signal, they could also be useful in the detection scenario. At present, we are investigating these issues.

Lastly, we would like to emphasise that the use of SVD in the context of tests of GR is not limited to Parametrised tests of PN theory.
It may be even more useful, for example, in the case of PPE  framework where the number of free parameters are much more due to the generality
of the waveform. It will be interesting in future to investigate the implications of SVD for PPE formalism.

{\it Note added} -- After this paper was finalised and circulated as a preprint, a related work by Brown et al~\cite{BrownEtal2012} appeared where they addressed detection problem of inspiralling spinning binaries. They discuss the construction of a template bank for spinning searches using a waveform where all of the PN terms in the phasing are treated as independent. They diagonalise the metric in the 8-dimensional parameter space to obtain a reduced two dimensional parameter space which they use for the template placement and the signal detection problem. They write down the PN phasing formula in terms of a dimensionless variable $x=f/f_0$ where $f_0$ is some arbitrary frequency scale. The new parameters we defined should correspond to $f_0=1$ Hz in their approach.

\section{Acknowledgements}
The authors thank B S Sathyaprakash  and E Berti for carefully reading the manuscript and for useful comments. The authors also thank P Ajith, S Bose, C Van Den Broeck, S V Dhurandhar, G Gonzalez, S Hild, B R Iyer, S Mitra, I Mandel, R Nayak, B F Schutz, A Sengupta, H Tagoshi for useful discussions and/or comments on the manuscript. KGA thanks IISER Thiruvananthapuram for hospitality during different stages of this work. This work is supported in part by AP's MPG-DST Max-Planck-India Partner Group grant. This document is a LIGO Document P1200075.
\section{References}
\bibliographystyle{iopart-num}
\bibliography{/home/arun/Dropbox/SVDproject/draft/Revised/ref-list}
\end{document}